\documentclass{article}
\usepackage{amsmath}
\usepackage[margin=1.in]{geometry}
\usepackage{amssymb}
\usepackage[square,sort,comma,numbers]{natbib}

\usepackage{lineno}
\usepackage{authblk}
\usepackage{tikz}
\usepackage{amsmath}
\usetikzlibrary{optics,calc,trees,shapes,decorations}

\title{True optical spacial derivatives for plasma density measurements}

\author[1,2]{P.-A. Gourdain\thanks{gourdain@pas.rochester.edu}}
\author[1,2]{I. N. Erez}
\author[1,2]{M. E. Evans}
\author[1,2]{H. R. Hasson}
\author[1]{J. Nagasako}
\author[1,2]{J. R. Young}
\author[1,2]{I. West-Abdallah}
\date{}

\affil[1]{Department of Physics and Astronomy, University of Rochester, Rochester NY 14627, USA}
\affil[2]{Laboratory for Laser Energetics, University of Rochester, Rochester NY 14627, USA}

\begin{document}
\maketitle

\begin{abstract}
This paper shows analytically and numerically that a vortex plate coupled to a neutral density filter can deliver a true optical derivative when placed at the focal plane of a $2f$ lens pair. This technique turns spatial variations in intensity into an intensity, which square root is the spatial derivative of the initial intensity variation. More surprisingly, it also turns any spatial variations \textit{in phase} into an intensity, which square root is the spatial derivative of the initial phase variation. Since the optical derivative drops the DC component of the signal, it is possible to measure the full electron plasma turbulence spectrum optically, without using any interferometer.
\end{abstract}
\section{Introduction}
The differentiation of a measured signal has always a task difficult to accomplish. This operation increases the noise to levels, with disastrous consequences on the subsequent analysis of a signal. This problem is common in scientific applications and, while the operation itself does not amplify the noise levels, the elimination of the signal average (also called "DC" value) decreases the signal-to-noise ratio. Yet, taking the space derivative of a signal can be useful. For instance, the derivative can enhance the contrast of an image. When light is used as a probe, the analog derivative operation becomes relatively straightforward, using optical systems to perform the derivative operation before the signal is recorded\cite{Hoffman:75}, leading to practical measurements such as mechanical shear \cite{Kulkarni2019}. In fact, almost any optical method has its own optical derivation technique, from schlieren imaging to \cite{Zakharin:04} to shearing interferometry\cite{Velghe2005}. Methods have been perfected over the years to yield really high sensitivity \cite{KEMAO2007,Legarda-Saenz2011,post1994moire}. Another advantage of using optics is the simplicity of doing a Fourier transform, which naturally appears on the focal plane of a converging lens. Any optical operation done on this plane is automatically done in Fourier space, and transferred back to the image when it is reconstructed on the other side of the lens. Using different filtering techniques ones can perform first order optical phase derivatives\cite{qian2003phase}, fractional derivatives\cite{Lancis:97,Tajahuerce97} and even two-dimensional first derivatives\cite{bracewell1986fourier}.

As manufacturing methods for optical devices reach sub-wavelength precision, new possibilities have come to light to form the derivative of an optical signal: vortex plates \cite{ORON2001325, oemrawsingh2004production}. By carving a spiral (stair-case like) cavity on the order of the light wavelength, one can generate vectorial beams\cite{Zhan:09,Galvez:12,boyd2016quantum,bouchard2016polarization}. When they are inserted on the focal plane of a thin lens, 
they can greatly improve the quality of the image from phase contrast imaging\cite{furhapter2005spiral,jesacher2005shadow}. However, a vortex plate coupled to a linearly varying neutral density filter placed on the focal plane of the $2f$ lens pair of Fig. \ref{fig:2f_setup} can yield a spatial derivative of extremely high quality. In this paper, we show how to construct an optical derivative using the mathematical foundations behind the optical Fourier transforms. We then use numerical simulations to show how this setup can be used to measure the electron density and turbulent fluctuations of a dense plasma. 
\section{True optical derivatives}
\subsection{Construction of a true optical derivative}
Lenses can decompose the image of an object into its spatial frequency components on the lens' focal plane
\begin{equation}\label{eq:fourier_theorem}
    U(f_x,f_y)=\int_{-\infty}^{+\infty}\int_{-\infty}^{+\infty}u(x,y)e^{i2\pi x f_x}e^{i2\pi y f_y}dxdy
\end{equation}
where the complex intensity $U(f_x,f_y)$ is now the Fourier transform of the object on the focal plane of the lens\cite{Goodman2017-eo}.

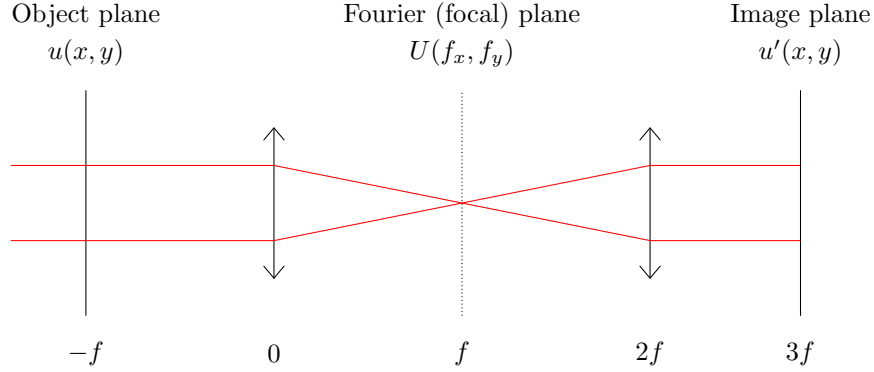
\begin{figure}[ht]
\centering
\begin{tikzpicture}[use optics]
\node[lens,focal length=2.5cm,object height=2cm] (L) at (0,0) {} ;
\node[lens,focal length=2.5cm,object height=2cm] (L1) at (5cm,0) {} ;
\coordinate (P) at (-3.5cm,0.5cm) ;
\coordinate (Q) at (-3.5cm,-0.5cm) ;
\coordinate (R) at (7cm,-0.5cm) ;
\coordinate (S) at (7cm,0.5cm) ;
\draw (-2.5cm,-1.5cm) -- (-2.5cm,1.5cm); 
\node at (-2.5cm,2.5cm) {Object plane};
\node at (-2.5cm,2cm) {$u(x,y)$};
\node at (-2.5cm,-2cm) {$-f$};
\node at (0cm,-2cm) {$0$};
\draw[densely dotted] (2.5cm,-1.5cm) -- (2.5cm,1.5cm); 
\node at (2.5cm,2.5cm) {Fourier (focal) plane};
\node at (2.5cm,2cm) {$U(f_x,f_y)$};
\node at (2.5cm,-2cm) {$f$};
\node at (5cm,-2cm) {$2f$};
\draw (7cm,-1.5cm) -- (7cm,1.5cm); 
\node at (7cm,2.5cm) {Image plane};
\node at (7cm,2cm) {$u'(x,y)$};
\node at (7cm,-2cm) {$3f$};
\def\toVerticalProjection#1#2#3{let \p{1} = #1, \p{2} = #2, \p{3} = #3 in
-- (\x{3},{\y{1}+(\y{2}-\y{1})/(\x{2}-\x{1})*(\x{3}-\x{1})})}
\draw[red] (P) -- ($(L.north)!(P)!(L.south)$) coordinate (Plens)
\toVerticalProjection{(Plens)}{(L.east focus)}{(L1)};
\draw[red] (Q) -- ($(L.north)!(Q)!(L.south)$) coordinate (Qlens)
\toVerticalProjection{(Qlens)}{(L.east focus)}{(L1)};
\draw[red] (R) -- ($(L1.north)!(R)!(L1.south)$) coordinate (Rlens)
\toVerticalProjection{(Rlens)}{(L1.east focus)}{(L1)};
\draw[red] (S) -- ($(L1.north)!(S)!(L1.south)$) coordinate (Slens)
\toVerticalProjection{(Slens)}{(L1.east focus)}{(L1)};
\end{tikzpicture}
\caption{Two-$f$ system uses two identical lenses of focal length $f$}
\label{fig:2f_setup}
\end{figure}

\subsubsection{Normalized spatial derivative using a vortex plate}
Following Ref. \citenum{Zhan:09}, using monochromatic coherent light, any image in the object plane $u(x,y)$ can be decomposed using the Fourier theorem as a discrete sum of sine and cosine functions
\begin{equation*}
u(x,y)=\sum_{n=-\infty}^{+\infty}A_n\cos(\textbf k_n\cdot\textbf r)+B_n\sin(\textbf k_n\cdot\textbf r),
\end{equation*}
where $\textbf k_n\cdot \textbf r=k_{x,n}x+k_{y,n}y$. While looking at the whole signal might be overwhelmingly complex to understand how the vortex plate operates, we can look at the components of $u(x,y)$ individually, labeled $u_{\textbf k}(x,y)=\cos(k_xx+k_yy)$. We drop the amplitudes $A_n$ and $B_n$ here for clarity. The corresponding signal on the lens focal plane of Fig. \ref{fig:2f_setup} is given by
\begin{equation}\label{eq:transform_cos}
    U_{\textbf k}(f_x,f_y)=\pi\delta\left(k_x/2\pi-f_x\right)\delta\left(k_y/2\pi-f_y\right)-\pi\delta\left(k_x/2\pi+f_x\right)\delta\left(k_y/2\pi+f_y\right),
\end{equation}
which evidently gives back $u'_{\textbf k}(x,y)=\cos(k_xx+k_yy)$ on the image plane located at $2 f$. However, when a vortex plate of charge $l=1$ is used on the Fourier plane, Eq. \ref{eq:transform_cos} turns into
\begin{equation}\label{eq:transform_cos_with_vortex}
    U_{\textbf k}(f_x,f_y)=\left[\pi\delta\left(\frac{k_x}{2\pi}-f_x\right)\delta\left(\frac{k_y}{2\pi}-f_y\right)-\pi\delta\left(\frac{k_x}{2\pi}+f_x\right)\delta\left(\frac{k_y}{2\pi}+f_y\right)\right]\exp\left[-i\arctan(f_x,f_y)\right].
\end{equation}
Note that the function $\arctan(f_x,f_y)$ gives the angle of the point $(f_x,f_y)$. Back on the image plane we now have
\begin{equation}\label{eq:image_cos_with_vortex}
    u'_{\textbf k}(x,y)=-\left(k_x+ik_y\right)\frac{\sin\left(k_xx+k_yy\right)}{\sqrt{k_x^2+k_y^2}}.
\end{equation}
At this point, we can compute the intensity $I_\textbf k$ of the mode wave number $\textbf k$ as
\begin{equation*}
    I_{\textbf k}=\left(k_x+ik_y\right)\frac{\sin\left(k_xx+k_yy\right)}{\sqrt{k_x^2+k_y^2}}\overline{\left(k_x+ik_y\right)\frac{\sin\left(k_xx+k_yy\right)}{\sqrt{k_x^2+k_y^2}}}
\end{equation*}
or $I_{\textbf k}=\sin^2\left(\textbf k\cdot\textbf r\right)$. We see that the square root of the intensity on the image plane located at $2f$ is \begin{equation}\label{eq:image_as_derivative}
    \sqrt {I_{\textbf k}}=\left|\sin\left(\textbf k \cdot \textbf r\right)\right|,
\end{equation}
which can be written as a \textit{normalized} derivative
\begin{equation}\label{eq:final_cos_with_vortex_plate}
    \sqrt {I_{\textbf k}}=\left|\frac{\partial\cos\left(\textbf k\cdot\textbf r\right)}{\partial\textbf k\cdot\textbf r}\right|,
\end{equation}
which simply corresponds to $|d\cos x/dx|$. Following the same reasoning for $u_{\textbf k}=\sin\left(\textbf k\cdot\textbf r\right)$ we get on the image plane
\begin{equation}\label{eq:final_sin_with_vortex_plate}
    \sqrt {I_{\textbf k}}=\left|\cos\left(\textbf k\cdot\textbf r\right)\right|=\left|\frac{\partial\sin\left(\textbf k\cdot\textbf r\right)}{\partial\textbf k\cdot\textbf r}\right|.
\end{equation}

\subsubsection{True optical derivative using a vortex plate and a neutral density filter}
However, Eq. \ref{eq:transform_cos_with_vortex} shows that it is relatively straightforward to recover a true derivative. If we add next to the vortex plate a neutral density filter, which transmission $T$ varies linearly with the radius as 
\begin{equation}
    T(f_x,f_y)=\sqrt{f_x^2+f_y^2},
    \label{eq:ND_profile}
\end{equation}
we now get on the Fourier plane
\begin{equation}\label{eq:true_optical_derivative}
    U_{\textbf k}(f_x,f_y)=\pi\left[\delta\left(\frac{k_x}{2\pi}-f_x\right)\delta\left(\frac{k_y}{2\pi}-f_y\right)-\delta\left(\frac{k_x}{2\pi}+f_x\right)\delta\left(\frac{k_y}{2\pi}+f_y\right)\right]\sqrt{f_x^2+f_y^2}\exp\left[-i\arctan(f_x,f_y)\right].
\end{equation}
Back to the image plane, Eq. \ref{eq:true_optical_derivative} yields
\begin{equation}\label{eq:image_cos_with_vortex_and_ND}
    u'_{\textbf k}(x,y)=-\left(k_x+ik_y\right)\sin\left(k_xx+k_yy\right).
\end{equation}
The square root of the intensity on the image plane is now
\begin{equation}\label{eq:final_cos_with_vortex_plate_with_ND}
    \sqrt {I_{\textbf k}}=\left|\textbf k \sin\left(\textbf k\cdot\textbf r\right)\right|=\left|\frac{\partial\cos\left(\textbf k\cdot\textbf r\right)}{\partial\textbf r}\right|.
\end{equation}
Using the same reasoning Eq. \ref{eq:final_cos_with_vortex_plate_with_ND} turns into
\begin{equation}\label{eq:final_sin_with_vortex_plate_with_ND}
    \sqrt {I_{\textbf k}}=\left|\textbf k\cos\left(\textbf k\cdot\textbf r\right)\right|=\left|\frac{\partial\sin\left(\textbf k\cdot\textbf r\right)}{\partial\textbf r}\right|
\end{equation}
when we look at the sine component of $u$.
With the help of the linearly varying neutral density filter, we obtained the actual spatial derivative of the signal $u_{\textbf k}(x,y)$ in the image plane. \begin{figure}[!htb]
    \centering
    a)\includegraphics[width=2.5in,trim={3.5cm 2.35cm 0 0}, clip]{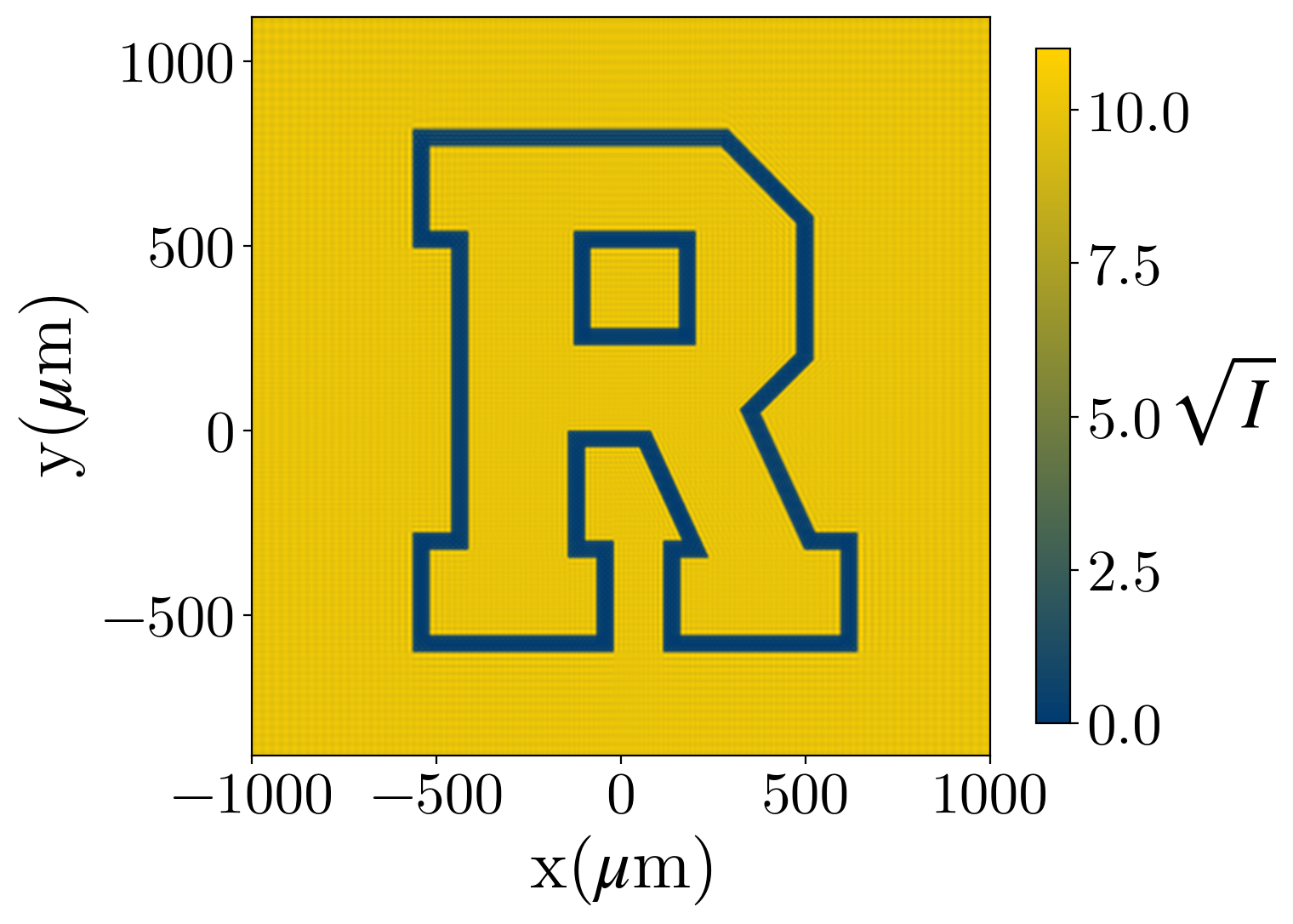}
    b)\includegraphics[width=2.5in,trim={3.5cm 2.35cm 0 0}, clip]{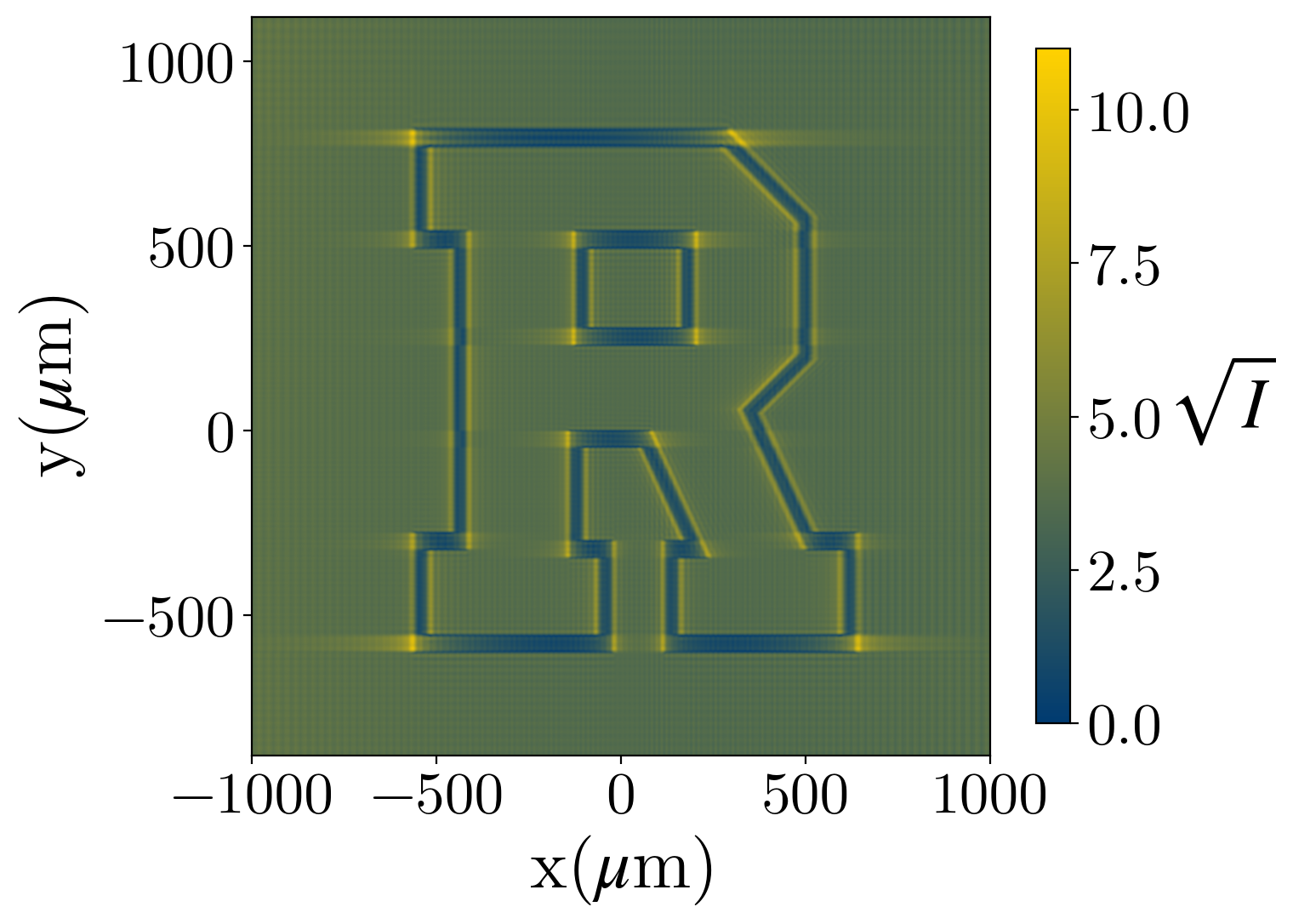}
    c)\includegraphics[width=2.5in,trim={3.5cm 2.35cm 0 0}, clip]{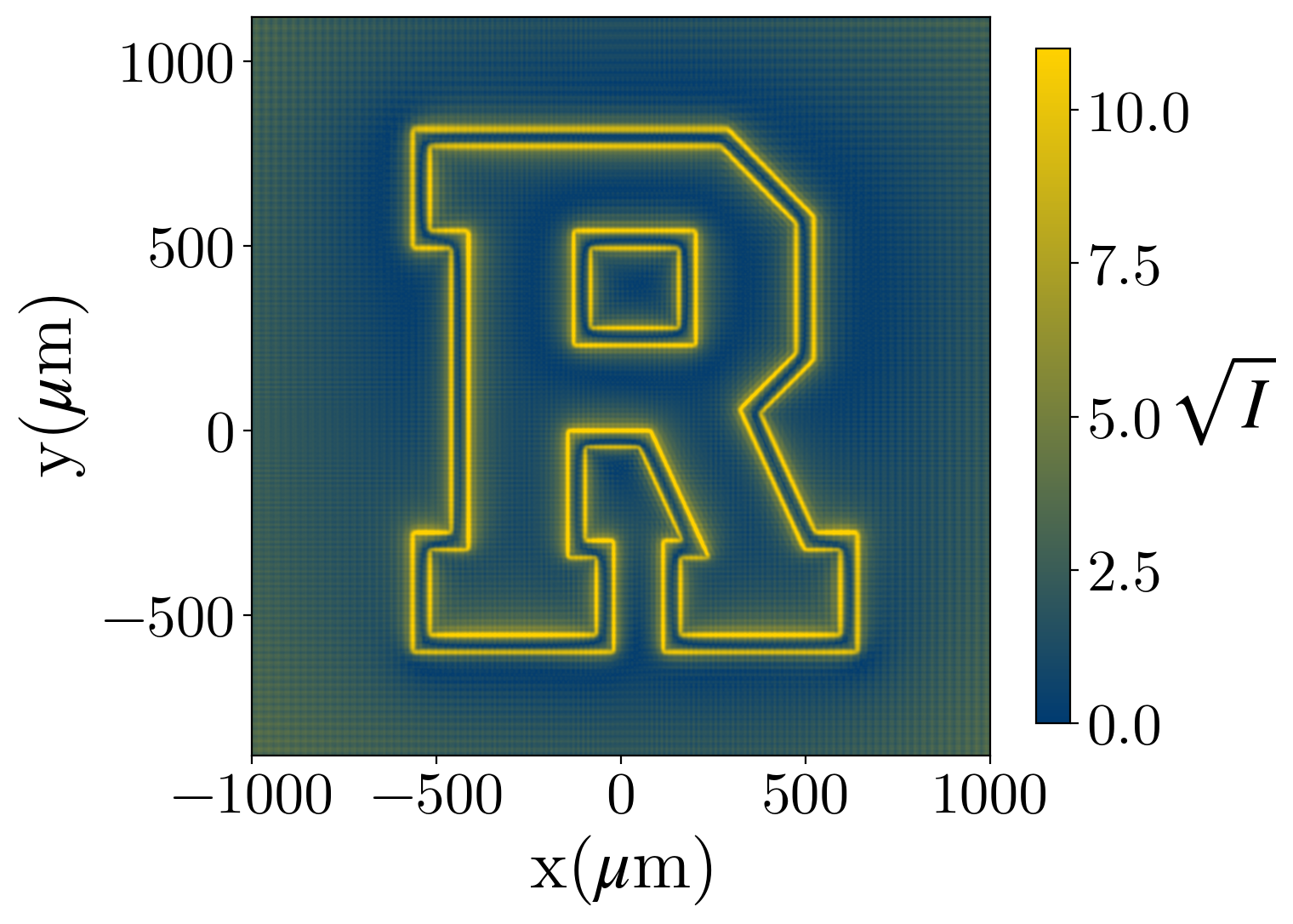}
    d)\includegraphics[width=2.5in,trim={3.5cm 2.35cm 0 0}, clip]{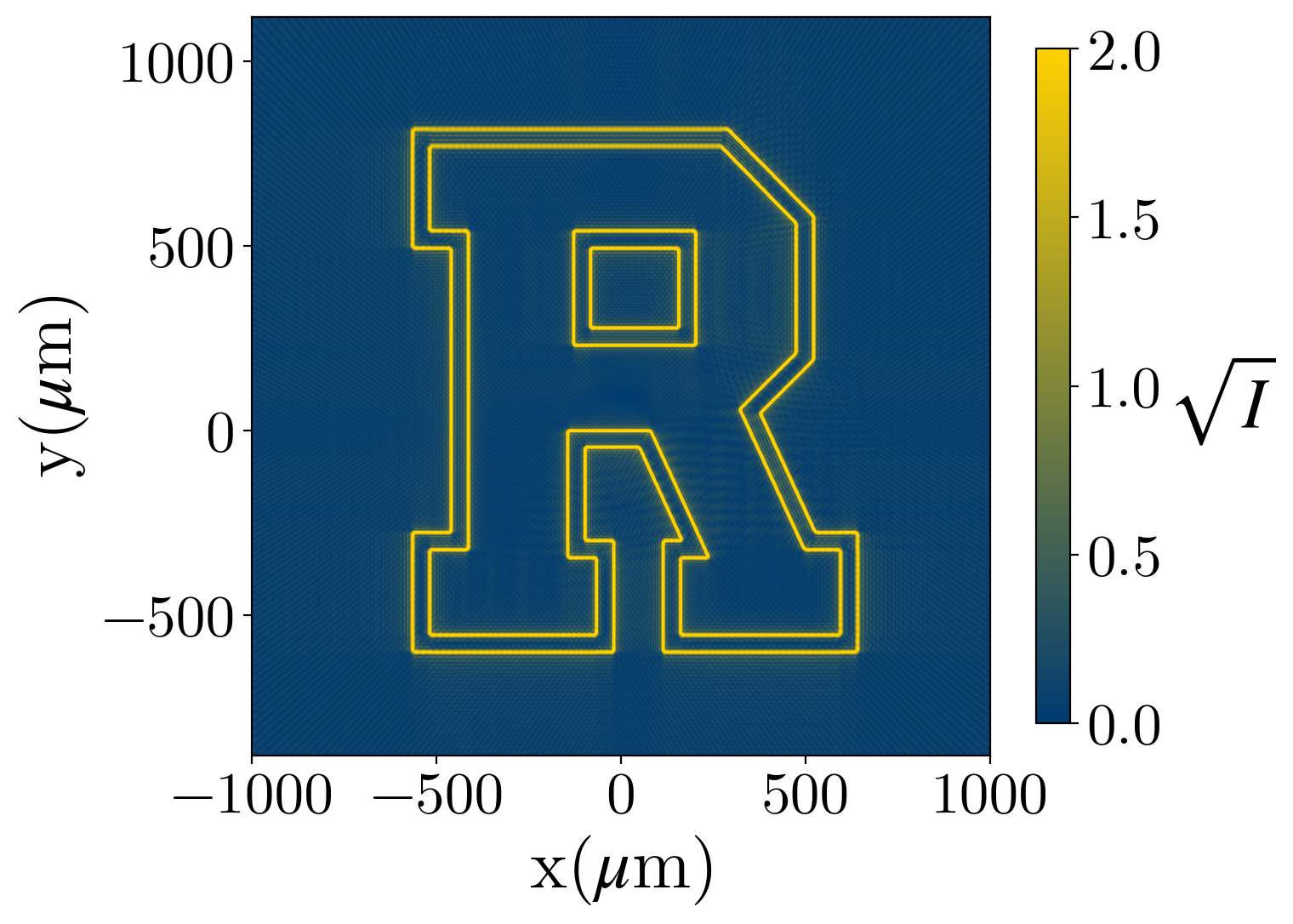}
     \caption{A mask of the spirit logo of the authors' university is placed at the object location shown in Fig. \ref{fig:2f_setup} leading to a) the bright-field image of this logo, located on the image plane. b) When a blade is inserted so its edge rests on the optical axis, we obtain a schlieren image. c) The image edges are sharper when a vortex plate is placed on the focal plane, yielding the normalized optical derivative. d) The image can be made much sharper by adding a neutral density filter, which produces the true optical derivative. All intensities are absolute.}
     \label{fig:spirit_logo}
\end{figure}

\subsection{Numerical validation}
In the rest of the paper, we use ray tracing\cite{brea2019} with Rayleigh-Sommerfeld diffraction \cite{shen2006fast} to compute the effect of each optical element on the intensity and phase of a Gaussian laser beam.  For practical reason we choose to work using a wavelength $\lambda=$532 nm.  The beam size $d$ is 8mm, the focal length of each lens is $f=10\textit{cm}$ and the total number of rays is 1024. We use numerical simulations here to highlight the difference between the normalized optical derivative and the true optical derivative. We also computed a schlieren image for comparison. The image of the backlit mask on the image plane is shown in Fig. \ref{fig:spirit_logo}-a. A well known method to enhance contrast uses schlieren imaging\cite{settles2017review}, shown in Fig. \ref{fig:spirit_logo}-b. We can greatly improve the contrast using vortex plate. As Fig. \ref{fig:spirit_logo}-c shows, the mask edges appear more clearly on the image plane. THe normalized derivation causes the "halo" surrounding the mask edges as large gradients have the same intensities as smaller ones. In other words, the derivative of large $k$ modes is similar to the derivative for low $k$ modes, leading to a spread of the derivative, regardless of its $k$ value. This result matches qualitatively Fig. 3 of Ref. \cite{jesacher2005shadow} using a spiral phase plate.  Finally, if we add the neutral density filter with profile given by Eq. \ref{eq:ND_profile} we finally get the true optical derivative, shown in Fig. \ref{fig:spirit_logo}-d. Based on Eqs. \ref{eq:final_cos_with_vortex_plate_with_ND} and \ref{eq:final_sin_with_vortex_plate_with_ND} we expect the intensity to be proportional to the characteristic wave number $k$ of the edge transition. 
\begin{figure}[!htb]
    \centering
    \includegraphics[width=4.75in]{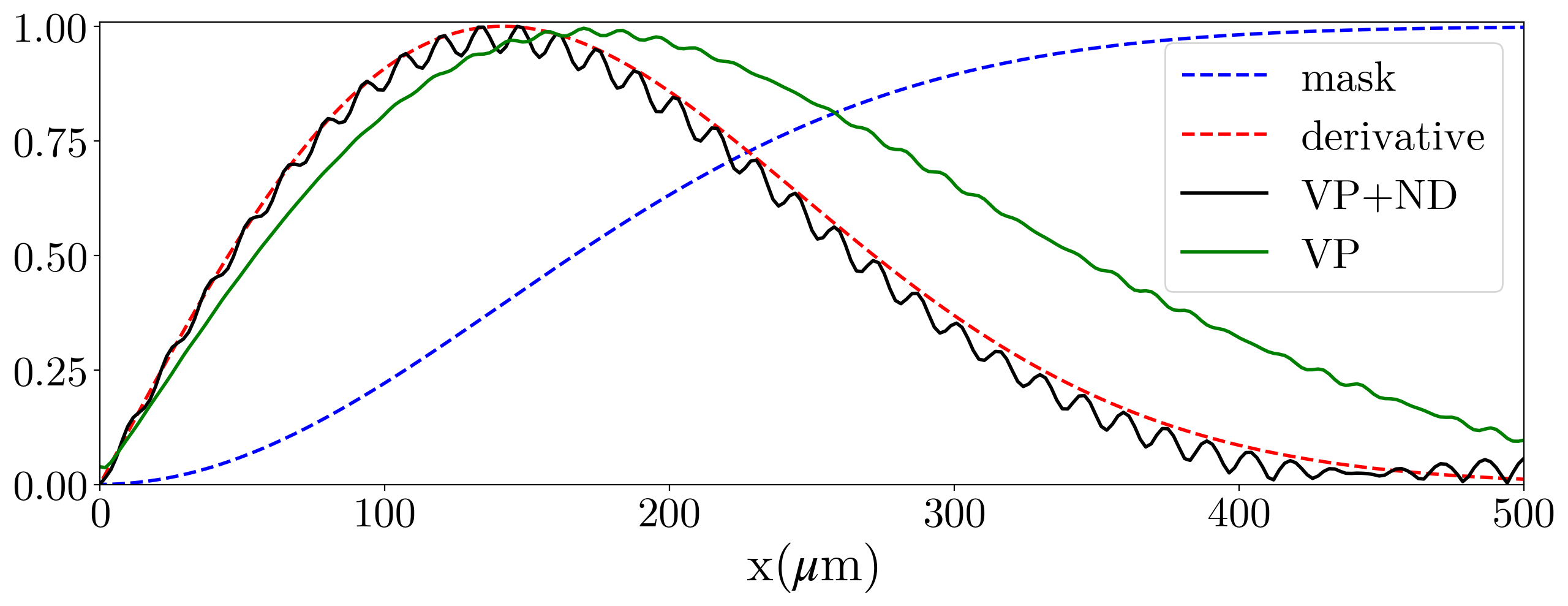} 
    \caption{The difference between the square root of the intensity from the vortex plate alone (VP) and the the vortex plate coupled to a neutral density filter (VP+ND) for a back-lit cylinder with a transmission profile given by $P(x)=1-\exp(-(x/\sigma)^2)$. The analytic derivative of the profile $P'(r)=\frac{2x}{\sigma^2}\exp(-(x/\sigma)^2) $ is also given for reference (Derivative). The mask transmission is also shown for reference. All line-outs have been normalized}
     \label{fig:check_derivative}
\end{figure}

We now use a mask that changes smoothly as
\begin{equation}
    P(x,y)=1-\exp\left(-\frac{x^2}{\sigma^2}\right),
    \label{eq:profile}
\end{equation}
to track numerically the value of the derivative and shown on Fig. \ref{fig:check_derivative}. This approach allows to check numerically the spatial dependence of the true optical derivative. Fig. \ref{fig:check_derivative} shows that the optical derivative matches the analytical derivative up to the simulation precision limited by the discretization inherent to the Rayleigh-Sommerfeld propagation scheme \cite{shen2006fast}. The separation between rays lead to some oscillations clearly visible on the line-out. These oscillations are due to the reduction in intensity, which decreases the signal to noise ratio of the computation. However, this is the maximum intensity for each ray that counts, and the intensity there always matches the analytical derivative. We can see that the normalized derivative using the vortex plate alone does not match as well the analytical derivative. Note that the RS propagation error is less pronounced for the vortex plate line-out (though still visible), since the actual intensity is an order of magnitude larger compared to the vortex plate coupled with the neutral density filter.

\section{Plasma density measurements}
We now pivot to the main topic of the paper. When a gas is ionized and turned into plasma with electron number density $N_e$, the index of refraction $n$ of the medium \cite{hutchinson2002} is given by
\begin{equation}
    n=\sqrt{1-\frac{N_e}{N_c}},
\end{equation}
where the critical electron density $N_c\left[\text{cm}^{-3}\right]\approx10^{21}\lambda[\mu\text{m}]^{-2}$. In this case, the phase change caused by electrons is given by
\begin{equation}
    \Delta\phi=\frac{\pi}{N_c\lambda}\int_LN_edl
\end{equation}
With simulations using green light at 532 nm, a phase shift of $2\pi$ corresponds to an areal electron density of $4\times10^{17}\text{cm}^{-2}$.
\subsection{Plasma density gradients}
\begin{figure}[!htb]
    \centering
    \includegraphics[width=4.75in]{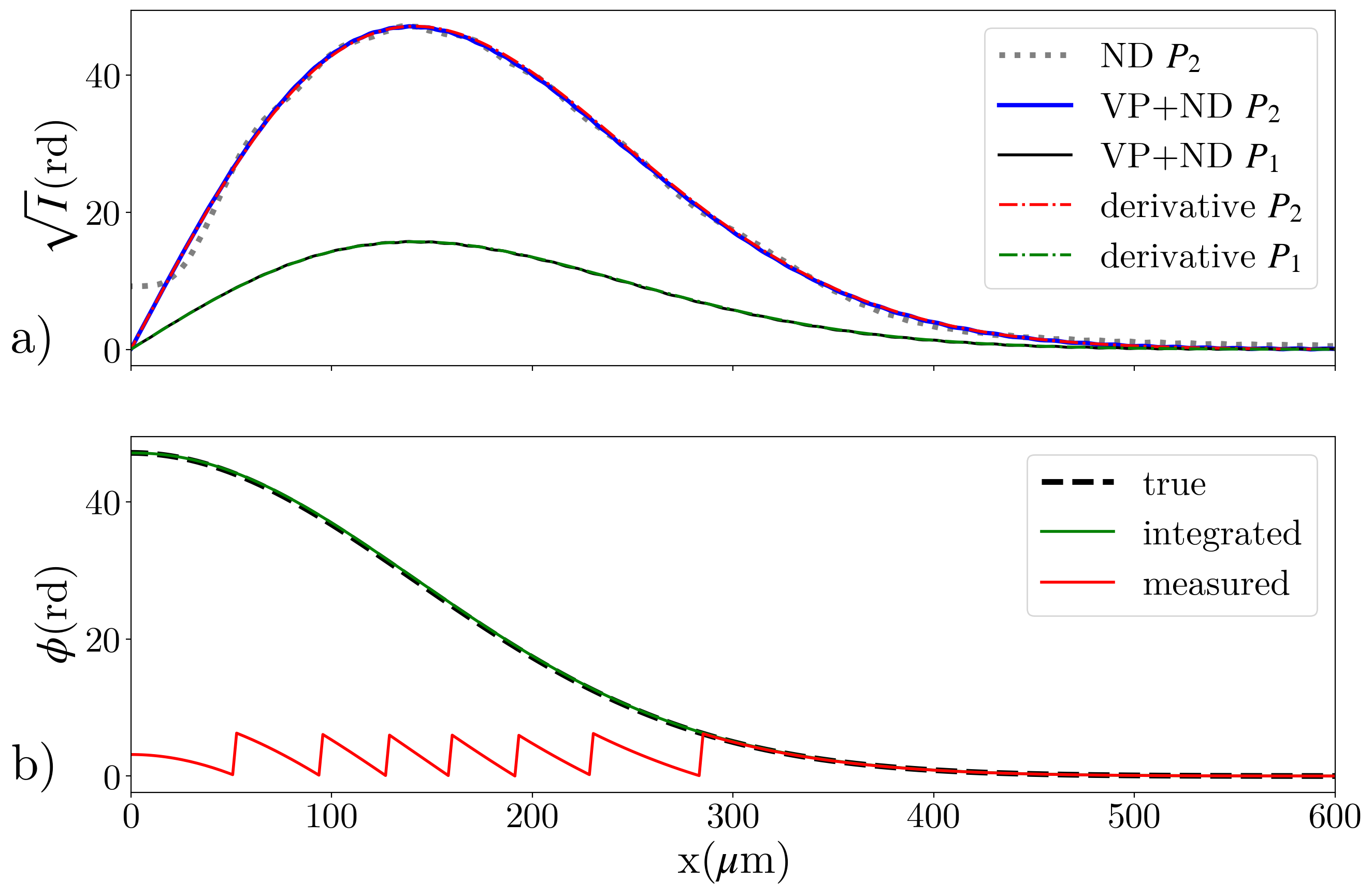} 
    \caption{a) The square root of the intensity on the optical axis along the positive part of the (horizontal) x-axis for two phase profiles $P_1$ and $P_2$. We scaled the peak of the square root of the intensity (VP+ND $P_1$) for the profile $P_1$ using the peak of $\partial P_1 /\partial x$ (derivative $P_1$) to calibrate the intensity. There is a clear agreement between the analytical phase derivative and the measured phase.  Keeping the calibration unchanged, we found excellent agreement between the square root of the intensity (VP+ND $P_2$) obtained for the phase profile $P_2$ and the analytical partial derivative $\partial P_2 /\partial x$ (derivative $P_2$). We also show the square root of the intensity obtained with the neutral density filter only (ND $P_2$). b) The integration of the square root of the intensity obtained for the phase profile $P_2$ (integrated) matches really well the analytical form of the phase profile $P_2$ (true). Without the optical derivative, only the wrapped phase can be measured (measured). }
     \label{fig:check_derivative_with_phase}
\end{figure}
If the plasma density is far from the cut-off density then laser interferometry can be used to measure the phase difference caused by free electrons inside the plasma. However this approach requires phase unwrapping\cite{ghiglia1998}, a task deceptively difficult. Here we propose to use the true optical derivative to measure density gradients. If the measurement is accurate enough, the density can be recovered by integrating the square root of the intensity measured on the image plane. So we trade phase unwrapping with absolute intensity measurement, which are prone to parasitic background light. Further, we only have access to the absolute value of the derivative, which makes integration challenging for plasma densities generated by complex plasma structures (e.g. Ref.  \cite{hasson2020design}).  

For the sake of simplicity, we suppose here that the plasma is completely transparent but has a density which line integrated phase varies as Eq. \ref{eq:profile}. We looked at two cases. The case where $P_1(x,y)=5\pi P(x,y)$ and the case $P_2(x,y)=15\pi P(x,y)$. The laser beam is then processed by the setup shown in Fig. \ref{fig:2f_setup}, leading to the results presented in Fig. \ref{fig:check_derivative_with_phase}.  After computing the intensity for the density profile corresponding to the distribution $P_1$, we scaled the square root to match the analytical derivative of $P_1$, in order to calibrate the simulation easily. Then we computed the intensity variation caused by the profile $P_2$ keeping the scaling unchanged. Fig. \ref{fig:check_derivative_with_phase}-a shows that the analytical derivatives are following exactly the derivatives obtained optically for both profile. We see that the optical and analytical derivatives virtually fall on top of each other for both $P_1$ and $P_2$. Note that when we use the neutral density filter without the vortex plate, we get an approximate optical derivative that does not match as well the analytical derivative on axis.

The integral of the optical derivative of $P_2$ is shown in Fig. \ref{fig:check_derivative_with_phase}-b. As expected, the integrated phase follows the true phase given by profile $P_2$. If we wanted to measure the true phase, we would get the phase wrapped modulo $2\pi$, also show in \ref{fig:check_derivative_with_phase}-b. Note that the proposed setup allows to measured variations in phase or in intensity in the exact same manner.

\subsection{Plasma turbulence}
\begin{figure}[!htb]
    \centering
    \includegraphics[height=1.6 in]{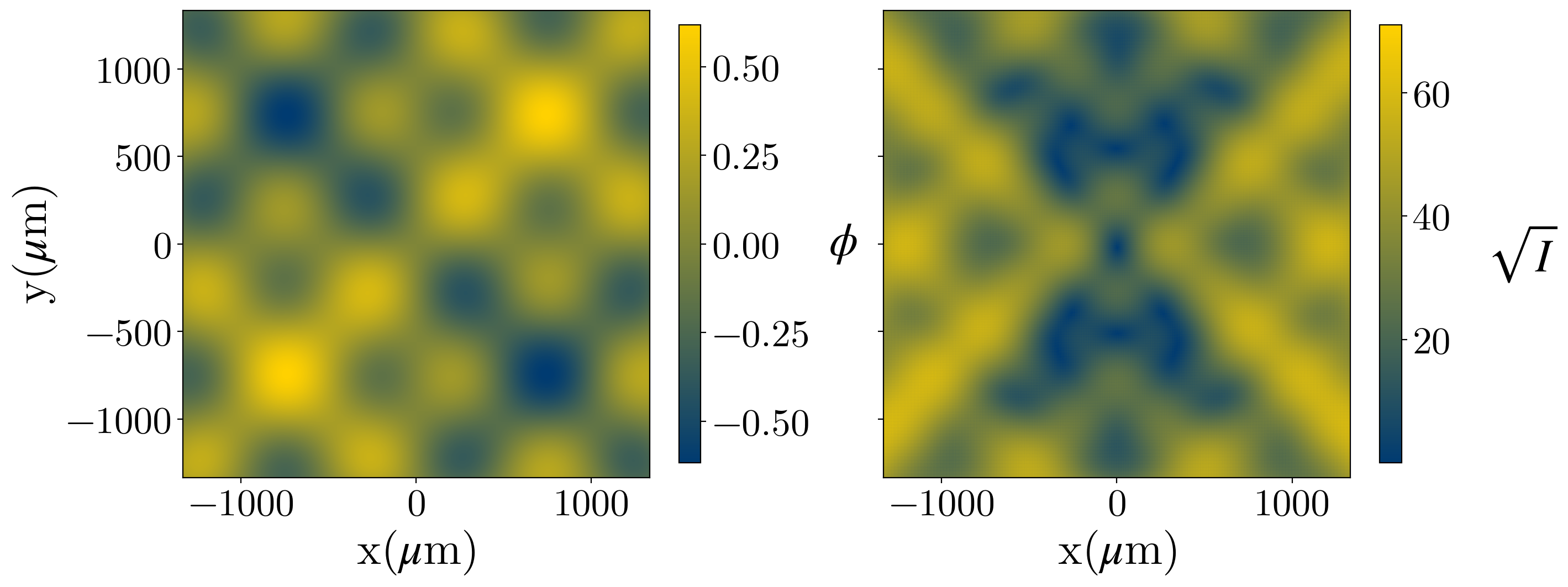}  \includegraphics[height=1.6 in,trim={20.5cm 0 0 0},clip]{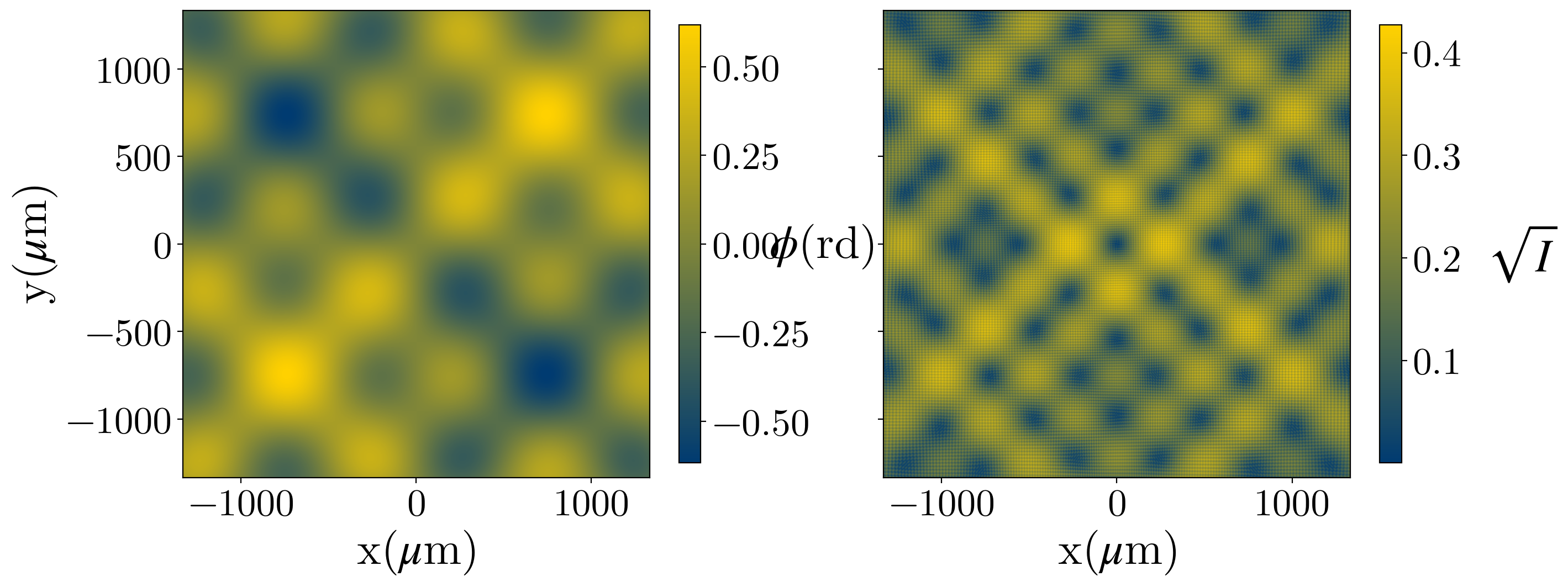}
    \caption{Phase fluctuations generated by Eq. \ref{eq:turb_equation} to simulate turbulence, with $k_1=3\pi/d$, $k_2=8\pi/d$, and $\gamma_1=\gamma_2=10$ (left).  The intensity is constant throughout the object plane. The square root of the intensity on the image plate for the vortex plate alone (center) and the vortex plate coupled with neutral density filter (right).}
     \label{fig:turbulence_derivative}
\end{figure}
Analog Fourier transforms using optical system to study the dominant modes inside an image preceded their digital counterpart by several decades. However, the removal of the zeroth order (i.e. DC) component \cite{Felstead:71} has always been necessary when one is interested in measuring all the modes\cite{lang1985}, especially for low wave numbers. Going back to Eqs. \ref{eq:final_cos_with_vortex_plate_with_ND} and \ref{eq:final_sin_with_vortex_plate_with_ND}, we clearly see that the optical derivative naturally cancels the constant light background (with wave number $k=0$). With the DC component removed from the Fourier plane, it is now possible to record low-k modes. As a proof of principle, we modelled plasma fluctuations following Ref. \cite{zweben2001plasma}, using the products of two sine functions,
\begin{equation}
    F(x,y)=\frac{\pi}{\gamma_1}\sin(k_1x)\sin(k_1y)+\frac{\pi}{\gamma_2}\sin(k_2x)\sin(k_2y).
    \label{eq:turb_equation}
\end{equation}
We included here two distinct modes $k_1$ and $k_2$. The corresponding density fluctuations are shown on the left in Fig. \ref{fig:turbulence_derivative} and the optical derivatives are on the right of Fig. \ref{fig:turbulence_derivative}. We took $\gamma_1=\gamma_2=10$, yielding a phase fluctuation areal density of $4\times10^{16}\text{cm}^{-2}$. When a vortex plate alone is used (center of Fig. \ref{fig:turbulence_derivative}), the periodicity of the image is lost. But when the neutral density filter is added, the derivative recovers the expected periodic structure. For our mode analysis, we now look at the Fourier plane where we can find the wave number of the turbulence directly. 

\begin{figure}[!htb]
    \centering
    \includegraphics[width=4.25 in]{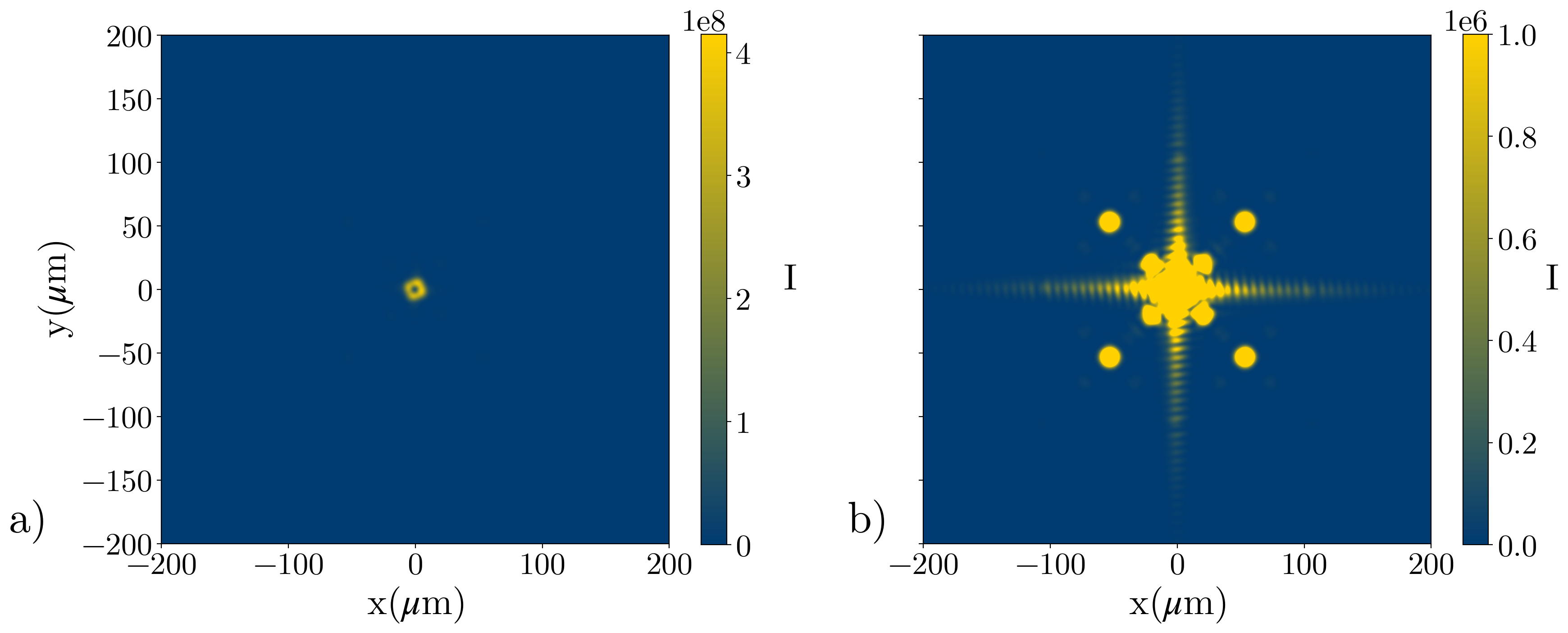} 
    \caption{The intensity on the Fourier (focal) plane, giving directly the spatial spectrum of the turbulence. a) Intensity plotted on the linear scale does not give much information on the turbulence spectrum. b) While the spectrum can be recovered on a saturated image, the DC component tends to wash out the low-k modes in practical implementations.  }
    \label{fig:k_turbulence_vp}
\end{figure}

\begin{figure}[!htb]
    \centering
    \includegraphics[width=4.25 in]{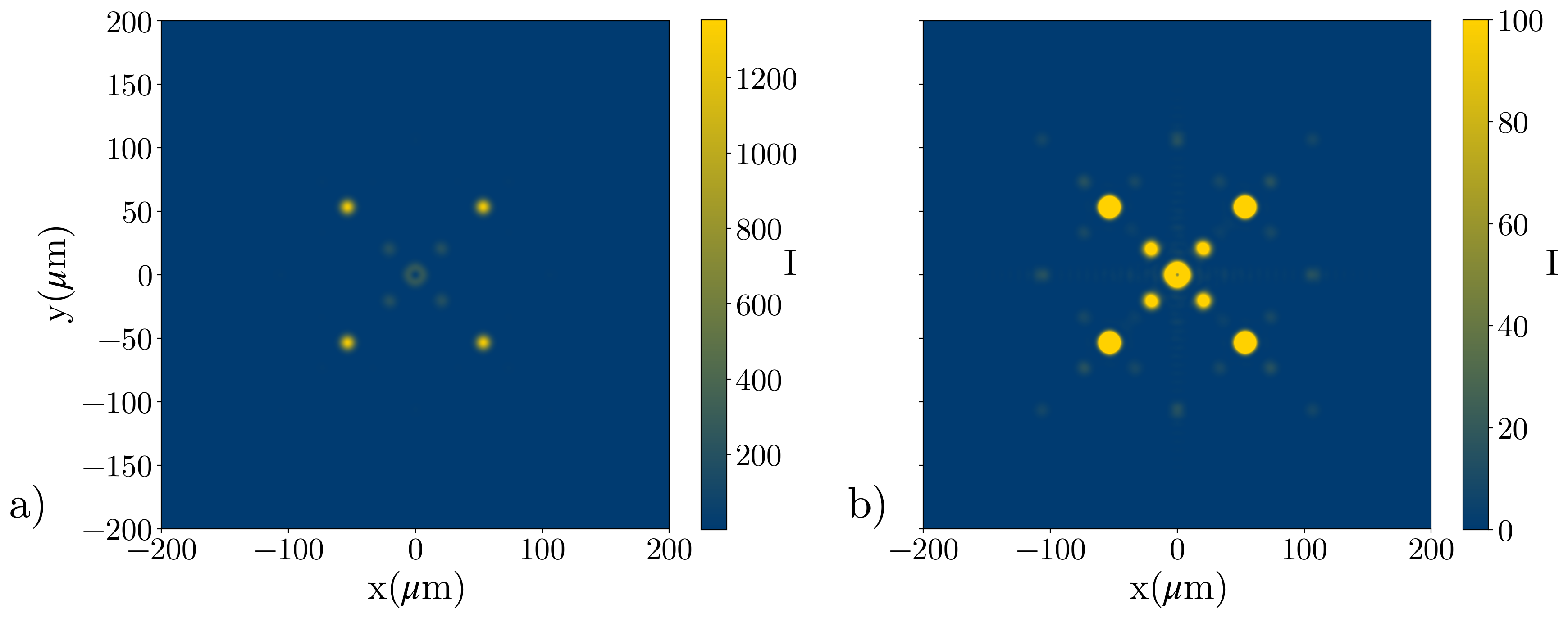} 
    \caption{The same density fluctuations as Fig. \ref{fig:k_turbulence_vp} but with the vortex plate coupled to the neutral density filter.  a) Even with the intensity on the linear scale we can see the different modes clearly. Due to the low DC level, even a saturated spectrum is fully usable.}
    \label{fig:k_turbulence_vpnd}
\end{figure}

Now starting with the vortex plate alone, Fig. \ref{fig:k_turbulence_vp}-a shows the DC component only when the intensity is plotted on the linear scale. Numerically, it is always possible to clip the color scale to regain some information on the dominant modes, even if the mode amplitude cannot be measured accurately (see Fig. \ref{fig:k_turbulence_vp}-b). Practically, the DC component tends to saturate a CCD detector, swamping the nearby pixels and limiting the measurement of low-k modes. By adding the neutral density filter, Fig. \ref{fig:k_turbulence_vpnd} shows the DC component of the signal has disappeared when taking the spatial derivative of the image and all the modes are now visible. Clipping the color only helps in seeing the mode harmonics. 
\section{Conclusions}
In this paper, we have shown that a \textit{2f} lens pair can be used to perform a true optical derivative by combining a vortex plate on the Fourier plane with a neutral density filter with linear radial dependence. It is relatively straightforward to obtain the derivative of two-dimensional intensity or phase profiles, though an absolute calibration is required to measure the plasma electron density directly, without an interferometer. However some uncertainties remain since we do not have access to the sign of the derivative, but only its absolute value. It is also possible to record directly the wave number of the turbulence by using the analog Fourier transform offered by the lens.

While this method is straightforward to implement numerically, some practical considerations still remain. Since we solely rely on intensity measurements, absolute calibration of intensity in the image plane is necessary, at least for measuring density. However, the absolute calibration is not required to measure the turbulence wave number. Ultimately, practical implementation will require a uniform laser beam, as speckling will be amplified by the derivative operation. Plasma light can be another source of noise that should be taken under consideration. So the beam needs to be filtered properly to generate a high quality backlighter.  

In conclusion, we were able to demonstrate analytically and numerically that a vortex plate coupled to a neutral density filter can deliver a true optical derivative when placed at the focal plane of a $2f$ lens pair. The system can be used to turn spatial variation in intensity into an intensity, which square root is the spatial derivative of the initial intensity variation. What is more surprising, the system also turns any spatial variation in phase into an intensity which square root is the spatial derivative of the initial phase variation, leading to a mean to measure phase without requiring interferometry. 

\section*{Acknowledgements}
This research was supported by the NSF CAREER Award PHY-1943939 and by the Laboratory for Laser Energetics Horton Fellowships.


\section*{References}
\bibliographystyle{ieeetr}
\bibliography{biblio}






\end{document}